\documentstyle[12pt,epsf]{article}

\catcode`\@=11

\def\cite{\@ifnextchar[{\@tempswatrue\@citex}{\@tempswafalse\@citex[]}}

\def\@citex[#1]#2{%
\if@filesw\immediate\write\@auxout{\string\citation{#2}}\fi
\leavevmode\unskip\ \@cite{\@collapse{#2}}{#1}}

\def\@bylinecite{%
\@ifnextchar[{\@tempswatrue\@CITEX}{\@tempswafalse\@CITEX[]}%
}

\def\@CITEX[#1]#2{%
\if@filesw\immediate\write\@auxout{\string\citation{#2}}\fi
\leavevmode\unskip$^{\scriptstyle\@CITE{\@collapse{#2}}{#1}}$}

\def\@cite#1#2{[{#1\if@tempswa , #2\fi}]} %
\def\@CITE#1#2{{#1\if@tempswa , #2\fi}} %

\def\@lbibitem[#1]#2{\item[\@BIBLABEL{#1}]\if@filesw
{\def\protect##1{\string ##1\space}\immediate
\write\@auxout{\string\bibcite{#2}{#1}}}\fi\ignorespaces}

\def\@biblabel#1{{[#1]}} %
\def\@BIBLABEL#1{$^{#1}\m@th$} %

\def\@collapse#1{%
{%
\let\@temp\relax
\@tempcntb\@MM
\def\@citea{}%
\@for \@citeb:=#1\do{%
\@ifundefined{b@\@citeb}%
{\@temp\@citea{\bf ?}%
\@tempcntb\@MM\let\@temp\relax
\@warning{Citation `\@citeb ' on page \thepage\space undefined}%
}%
{\@tempcnta\@tempcntb \advance\@tempcnta\@ne
\edef\MyTemp{\csname b@\@citeb\endcsname}%
\def\@tempa{\@temptokena=\bgroup}%
\afterassignment\@tempa %
\@tempcntb=0\MyTemp\relax}%
\ifnum\@tempcntb=0\relax%
\@tempcntb=\@MM
\@citea\MyTemp
\let\@temp = \relax
\else %
\edef\@tempd{\number\@tempcntb}%
\ifnum\@tempcnta=\@tempcntb %
\ifx\@temp\relax %
\edef\@temp{\@citea\@tempd}%
\else
\edef\@temp{\hbox{--}\@tempd}%
\fi
\else %
\@temp\@citea\@tempd
\let\@temp\relax
\fi
\fi
}%
\def\@citea{,}%
}%
\@temp %
}%
}%
\catcode`\@=12

\makeatletter
\@addtoreset{equation}{section}
\makeatother

\def\dj{\hbox{d\kern-0.347em \vrule width 0.3em height 1.252ex depth
-1.21ex \kern 0.051em}}
\def\Tr{{\rm Tr\,}}
\def\e{{\rm e}}
\def\d{{\rm d}}
\def\CO{{\cal O}}
\begin{document}
\title
{Correlation Functions in Matrix Models \\[0.3cm]
Modified by Wormhole Terms\\[1.0cm]}

\author{
J. L. F. Barb\'on, \
K. Demeterf\kern 1.05em\kern -1em i\thanks{On leave of absence from
the Ru{\dj}er Bo{\v s}kovi{\'c} Institute, Zagreb, Croatia}\,, \
I.~R. Klebanov, \
C. Schmidhuber\thanks {On leave of absence from the Institute of
Theoretical Physics, University of Bern, Switzerland} \\[0.2cm]
Joseph Henry Laboratories\\
Princeton University\\
Princeton, New Jersey 08544, USA
}

\date{}
\maketitle
\setcounter{page}{0} \pagestyle{empty}
\thispagestyle{empty}
\vskip 1.0in
\begin{abstract}
We calculate correlation functions in matrix models modified by
trace-squared terms.
First we study scaling operators in modified one-matrix models and
find that their
correlation functions satisfy modified Virasoro constraints.
Then we turn to dressed order parameters in minimal models
and show that their correlators satisfy Goulian-Li formulae
continued to negative Liouville dressing exponents.
Our calculations provide additional support for the idea that the
modified matrix models contain operators with the negative branch of
gravitational dressing.
\end{abstract}

\vfill
\begin{flushleft}
PUPT-1517, \ \
hep-th/9501058 \\
January 1995
\end{flushleft}
\newpage \pagestyle{plain}

\section{Introduction}

In recent literature some effort has been devoted to studying the
large $N$ matrix models modified by trace-squared
terms~\cite{DDSW,ABC,Ko,SuTs,GuKl,Kl,KlHa,Sh}.
In the one-matrix models, for instance, one may add terms like
$ g(\Tr\Phi^4)^2$ to the conventional matrix potential of the
form $N\Tr V(\Phi)$.

Discretized random surfaces appear in the Feynman graph expansion of
the matrix models, and the vertex corresponding to the
trace-squared term may be thought of as an identification between
a pair of ordinary plaquettes.
If the two plaquettes belong to
otherwise disconnected random surfaces, then this identification
introduces a
tiny ``neck'' into the geometry; if they belong to the same connected
component, then the identification attaches a
microscopic handle.
These two types of effects are familiar consequences of the
Euclidean wormholes, which were widely explored in the four-dimensional
quantum gravity. Thus, trace-squared terms are a convenient trick for
adding wormholes to the random surface geometry, with the correct
combinatorics emerging automatically. Actually, wormholes are known
to be abundant already in conventional random surface theories
with no trace-squared terms~\cite{AgMi,JaMa,KKMW}. Therefore,
turning on the coupling $g$ does not introduce
any new types of geometry into the path integral, but simply increases the
weight of microscopic wormhole configurations.
For example, in the sum over surfaces of genus zero there are geometries
corresponding to trees of smooth spheres glued pairwise at single
plaquettes, and increasing $g$ enhances their weight.
It is not surprising, therefore, that a small increase in $g$ does not change
the universal properties of the model. If, however, $g$ is fine-tuned to a
finite positive value $g_t$, then the universality class
of the large area behavior changes.
For pure gravity ($c=0$) the
string susceptibility exponent jumps from $-1/2$
for $g<g_t$ to $1/3$ for $g=g_t$.
This is the simplest example of a matrix model where new critical behavior
occurs due to fine-tuned wormhole weights.

Further work has revealed that, more generally,
as the trace-squared coupling is increased to a
critical value $g_t$, the string susceptibility exponent jumps from
some negative value $\gamma$, found in a conventional matrix model, to
a positive value
\begin{equation} \label{newgamma}
\bar{\gamma}={\gamma\over\gamma-1}\ .
\end{equation}
Essentially equivalent results have been obtained without using matrix models,
on the basis of direct combinatorial analysis~\cite{Du}. For a long time the
positive values of string susceptibility exponent seemed very puzzling.
Recently, however, a simple continuum explanation of these critical
behaviors was proposed in ref.~\cite{Kl}.

For all the conventional matrix models describing $(p, q)$ minimal
models coupled to gravity, the correct scaling follows from the
Liouville action of the form
\begin{eqnarray} \label{usint}
S_L&=&{1\over 8\pi}\int \d^2\sigma \, \left( (\partial_\mu \phi)^2-
Q\hat R \phi +t O_{\min} \e^{\alpha_+\phi} \right)\ ,
\nonumber \\
\alpha_+ &=&
{1\over 2\sqrt 3} \left(\sqrt {1-c+24h_{\min}}-
\sqrt{25-c}\right)=-{p+q-1\over \sqrt{2pq}}\ ,
\end{eqnarray}
where $O_{\min}$ is the matter primary field of the lowest dimension.
A simple calculation reveals that the
string susceptibility exponent is given by
\begin{equation} \label{geng}
\gamma= 2+ {Q\over\alpha_+}\ ,
\end{equation}
where
$ Q=\sqrt{25-c\over 3} $.
In ref.~\cite{Kl} it was argued that the effect of fine-tuning the
touching interaction is to replace the positively dressed
Liouville potential by the negatively dressed one,
\begin{eqnarray} \label{modpot}
S_L &=& {1\over 8\pi}\int \d^2\sigma\, \left( (\partial_\mu \phi)^2-
Q\hat R \phi + \bar t O_{\min} \e^{\alpha_-\phi}\right)\ ,
\nonumber \\
\alpha_- &=&
-{1\over 2\sqrt 3}\left(\sqrt {1-c+24h_{\min}}+
\sqrt{25-c}\right)=-{p+q+1\over \sqrt{2pq}} \ .
\end{eqnarray}
Now the string susceptibility exponent is found to be
\begin{equation} \label{newg}
\bar{\gamma}= 2+ {Q\over\alpha_-}={\gamma\over \gamma-1}\ ,
\end{equation}
in agreement with the matrix model results.
The intuitive reason for the change of the branch of gravitational
dressing is that the microscopic wormholes alter the ultraviolet
(large $\phi$) behavior of the theory. A priori, there are
two independent solutions to the string equations of motion,
corresponding to the two choices of dressing. The linear combination
that appears in the theory is selected by the boundary conditions, and
by fine-tuning the boundary
conditions at large $\phi$ we have changed the string background.

A more complete understanding of the modified matrix models was recently
found in ref.~\cite{KlHa}.
As the trace-squared coupling is set to $g_t$ one finds the following
non-perturbative relation
\begin{equation} \label{double}
\e^{\bar F (\bar t; t_1, t_2, \ldots, t_n)} =
\int_{-\infty}^\infty \d t\,
\e^{t \bar t + F(t; t_1, t_2, \ldots, t_n)}\ ,
\end{equation}
where $\bar F$ and $F$ are the universal parts of the modified and
conventional free energies, respectively. As indicated in eqs.~(\ref{usint})
and (\ref{modpot}), $t$ is the lowest dimension coupling in the
conventional action and $\bar t$ is the corresponding coupling in the modified
action; $t_i$ are coupling constants corresponding to other operators.
Evaluating the integral in (\ref{double}) in saddle-point expansion, we may
generate the genus expansion of $\bar F$ in terms of the known
genus expansion of $F$. This gives a concise prescription for calculating
universal correlation functions in modified matrix models, which will
be used extensively in this paper.

A further result of ref.~\cite{KlHa} is that, by changing the type of
trace-squared terms,
it is possible to introduce integrations over other couplings $t_i$.
For example, if $t_1$ is the coupling constant for a matrix model
scaling operator $O_1$, then by adding $g_1 O_1^2$ to the matrix action and
fine-tuning $g_1$, we obtain a model where
\begin{equation} \label{newdouble}
\e^{\bar F (t; \bar t_1, t_2, \ldots, t_n)} =
\int_{-\infty}^\infty \d t_1 \,
\e^{t_1 \bar t_1 + F(t; t_1, t_2, \ldots, t_n)}\ .
\end{equation}
The dependence of $\bar F$ on $\bar t_1$ indicates that the gravitational
dimension of $O_1$ has changed from its conventional value $d$ to
\begin{equation} \label{newgd}
\bar d=\gamma-d
\end{equation}
(the string susceptibility exponent remains
unchanged). Remarkably, this change of
dimension is again reproduced in Liouville theory by a mere
change of the branch of gravitational dressing. Namely, if the operator
is dressed by $\e^{\beta_\pm\phi}$ then
$$ d= 1- {\beta_+\over \alpha_+}, \qquad\qquad
\bar d= 1- {\beta_-\over \alpha_+}\ .
$$
The relation (\ref{newgd}) follows from $\beta_++\beta_-= -Q$.

By applying integral transformations (\ref{double}) and
(\ref{newdouble})
sequentially to any subset of coupling constants, we can construct
a matrix model where the scaling dimensions of all the corresponding
operators correspond to picking the {\it negative} branch of Liouville
dressing.
It appears that we have found a whole new class of matrix models which
serve as exact solutions of a new class of Liouville theories, those involving
some number of negatively dressed operators. The purpose of this paper
is to use the new matrix models to extract some information about
the negatively dressed operators and to compare it, if possible, with
direct continuum arguments.

In section 2 we analyze the correlation functions of macroscopic loop
operators $\bar{w}(\ell)$ in modified one-matrix models.
Using eq.~(\ref{double}) to implement
the change of dressing of $O_{\min}$, we calculate the loop correlators
on a sphere and find agreement with the results of refs.~\cite{ABC,Ko}.
Correlation functions of the scaling operators $\bar\sigma_n$ may be read off
from the expansion of $\bar{w}(\ell)$ in powers of $\ell$.
In addition to the basic
operator $\bar\sigma_0$ which by definition has gravitational dimension 0,
we find operators $\bar\sigma_n$, $n>0$, of dimensions $\bar\gamma (n+1)$.
Remarkably, there is no operator of dimension $\bar\gamma$ which in the
Liouville language would correspond to $O_{\min} \e^{\alpha_+\phi}$.
This is not a coincidence: operator $O_{\min} \e^{\alpha_-\phi}$ already
appears in the theory, and a simultaneous appearance of
$O_{\min} \e^{\alpha_+\phi}$ would signify a doubling of the spectrum that
is unacceptable on general grounds.\footnote{As we remarked before, in any
theory we expect the boundary conditions to single out unique gravitational
dressing.}

The dependence of $\bar F$ on $t_n$, the coupling constants for
$\bar\sigma_n$, can be found from
the general formula (\ref{double}) (an explicit derivation of this fact
for this particular system appears in Appendix A). From the fact that
$\e^F$ obeys the Virasoro constraints, it follows that $\e^{\bar F}$ obeys
modified Virasoro constraints, obtained from the conventional ones by replacing
\begin{equation} \label{bogo}
t \rightarrow {\partial\over \partial\bar t}\ ,
\qquad\qquad {\partial\over \partial t} \rightarrow -\bar t\ .
\end{equation}
We discuss the Virasoro constraints and the recursion relations they imply
among the correlation functions in section 3.

In section 4 we check formula (\ref{newdouble}) for transformation of
scaling operators in the context of one-matrix models. For this formula
to apply to gravitational descendants, a class of leading analytic
terms must vanish in their conventional correlation functions.
We check this vanishing for some specific examples.

In section 5 we use eq.~(\ref{double}) and formulae for the spherical
two- and three- point correlators of positively dressed $c<1$ order parameters
to calculate correlators involving the negatively dressed order parameters.
We find simple results which agree with a straightforward
analytic continuation in the Goulian-Li formulae~\cite{GoLi}.
As a further check,
we carry out similar analysis for the $c=1$ theory up to the four-point
function which is, fortunately, known explicitly.

\section{Scaling Operators and Macroscopic Loops}

In this section we investigate the correlation functions of scaling
operators in the modified multicritical one-matrix
models~\cite{ABC,Ko},
\begin{equation} \label{pfun}
Z_k = \int \,{\cal D}\Phi\, \e^{-N\left(\Tr V_k (\Phi) + (c_2 - \lambda)
\Tr \Phi^4 -{g\over 2N} (\Tr \Phi^4)^2 \right)} .
\end{equation}
The critical potential of the $k$th model with $g=0$ is
\begin{equation} \label{kpot}
V_k (\Phi) = \sum_{i=1}^{k} c_i \, \Phi^{2i}\ ,
\end{equation}
where $c_i$ have been determined in ref.~\cite{oldMM}.

The correlation functions of the scaling operators for $g=0$ have
been studied in ref.~\cite{oldCF}. The scaling operators are given by
linear combinations of traces of powers of $\Phi$,
\begin{equation} \label{so}
\sigma_{n}=\sum_{i=1}^{n+1} g_{i}^{(n)} \Tr\Phi^{2i}\ ,\qquad
n=0,1,2,\dots \ .
\end{equation}
The coefficients $g_{i}^{(n)}$ are chosen so that the correlation
functions of $\sigma_{n}$'s on the sphere scale as~\cite{oldCF}:
\begin{equation} \label{cf}
\langle\sigma_{1}\dots\sigma_{p}\rangle\sim t^{2-\gamma+
\sum_{n}(d_{n}-1)}\ ,
\end{equation}
where $t\sim(c_2-\lambda)N^{2/(2-\gamma)}$, and
$d_n=-n\gamma$ is the gravitational dimension of the operator
$\sigma_n$.

This analysis can be extended to the $g\neq 0$ case. For the
fine-tuned $g=g_t$, in addition to the dimension $0$ operator,
$\bar\sigma_0=\Tr\Phi^2$, one finds a set of ``modified'' scaling
operators
\begin{equation} \label{mso}
\bar\sigma_n = \sum_{i=1}^{n+1}\bar g_{i}^{(n)} \Tr \Phi^{2i}\ ,\qquad
n=1,2,3,\dots \ ,
\end{equation}
with gravitational dimensions $\bar d_n=(n+1)\bar\gamma$.
The correlation functions of these new operators on the sphere are
given by
\begin{equation} \label{mcf}
\langle\bar\sigma_{1}\dots\bar\sigma_{p}\rangle\sim
\bar t^{\,2-\bar\gamma+\sum_{n}(\bar d_{n}-1)}\ ,
\end{equation}
where $\bar t\sim (\lambda_c-\lambda)N^{2/(2-\bar\gamma)}$.
There are two interesting features of these results we want to
emphasize. First, the coefficients $\bar g_{i}^{(n)}$ in~(\ref{mso})
are equal to the coefficients $g_{i}^{(n)}$ in~(\ref{so}) for all
$i$ except $i=2$. A general expression for $\bar g_{2}^{(n)}$ in the
$k$th multicritical model is given in Appendix A. Second, there is no
operator of dimension $\bar\gamma$. This provides additional argument
in favor of Liouville interpretation of the modified matrix models~\cite{Kl}.
In Liouville theory such an operator would correspond to
$O_{\min} \e^{\alpha_+\phi}$ which is not acceptable. Namely, by
choosing the negative branch of dressing we already have operator
$O_{\min} \e^{\alpha_-\phi}$ in the theory, and the existence of
the operator $O_{\min} \e^{\alpha_+\phi}$ would mean a doubling of
the spectrum.

In Appendix A we carefully derive eq.~(\ref{double}) which
generalizes the relation between the universal parts of modified
and conventional free energies in the presence of perturbations
of the multicritical potentials.
This result allows us to calculate arbitrary correlation function of
$\bar\sigma_n$'s for any genus once the corresponding correlation
functions of $\sigma_n$'s are known.

It is actually more convenient to consider the macroscopic loop operator,
$w(\ell)$, which contains the complete information about correlation
functions of scaling operators $\sigma_{n}$'s in a compact form,
\begin{equation} \label{wexp}
w(\ell)= \sum_{n=0}^{\infty}\ {\ell^{n+1/2} \over \Gamma(n+3/2)}\
\sigma_{n}\ .
\end{equation}

Let us assume that $t_{i}$'s in eq.~(\ref{double}) are couplings to
macroscopic loops $w(\ell_{i})$, and denote them by $r_i$.
Then we have
\begin{equation} \label{nloop}
\langle w(\ell_1) \dots w(\ell_p)\rangle =
{\partial^p\over\partial r_1 \dots \partial r_p}\, F(t;r_i)\ ,
\end{equation}
and
\begin{equation} \label{mnloop}
\langle\bar w(\ell_1) \dots \bar w(\ell_p)\rangle =
{\partial^p\over\partial r_1 \dots \partial r_p}\, \bar F(\bar t;r_i)\ .
\end{equation}
We now show how correlation functions in the modified models~(\ref{mnloop})
can be systematically calculated in terms of those in the
conventional $c<1$ models~(\ref{nloop}).

Let us first consider the correlators on the sphere. In this case
eq.~(\ref{double}) can be written as Legendre transform
\begin{equation} \label{Legendre}
\bar F(\bar t;r_i)=t\bar t + F(t;r_i) \ ,
\end{equation}
where the r.h.s. is evaluated at the saddle-point $t_{s}(\bar t\,)$,
given by the solution of equation
\begin{equation} \label{saddle}
\bar t= -{\partial F(t;r_i)\over\partial t} \ .
\end{equation}
By simply taking derivatives of~(\ref{Legendre}), and using the chain rule,
one finds for the one-loop,
\begin{equation} \label{m1l}
\langle\bar w(\ell)\rangle_{0}(\bar t\,) =
\langle w(\ell)\rangle_{0}\bigg\vert_{t=t_{s}(\bar t\,)} \ ,
\end{equation}
two-loops,
\begin{equation} \label{m2l}
\langle\bar w(\ell_1) \bar w(\ell_2)\rangle_{0}(\bar t\,) =
\left\{\langle w(\ell_1) w(\ell_2)\rangle_{0} -
{\langle w(\ell_1) P\rangle_{0} \langle P w(\ell_2)\rangle_{0} \over
\langle PP \rangle_{0}} \right\}\Bigg\vert_{t=t(\bar t\,)} \ ,
\end{equation}
three-loops,
\begin{eqnarray} \label{m3l}
&&\langle\bar w(\ell_1)\bar w(\ell_2)\bar w(\ell_3)\rangle_{0}(\bar t\,)=
\Biggl\{
\langle w(\ell_1) w(\ell_2) w(\ell_3)\rangle_{0} -
{\langle w(\ell_1) w(\ell_2) P\rangle_{0}\langle P w(\ell_3)\rangle_{0}\over
\langle PP \rangle_{0}}
\nonumber \\
&&\qquad - \
{\langle w(\ell_1) w(\ell_3) P\rangle_{0}\langle P w(\ell_2)\rangle_{0}\over
\langle PP \rangle_{0}} -
{\langle w(\ell_2) w(\ell_3) P\rangle_{0}\langle P w(\ell_1)\rangle_{0}\over
\langle PP \rangle_{0}}
\nonumber \\
&&\qquad+ \
{\langle w(\ell_1) PP\rangle_{0}\langle P w(\ell_2)\rangle_{0}
\langle P w(\ell_3)\rangle_{0}\over (\langle PP \rangle_{0})^2 } +
{\langle w(\ell_2) PP\rangle_{0}\langle P w(\ell_3)\rangle_{0}
\langle P w(\ell_1)\rangle_{0}\over (\langle PP \rangle_{0})^2 }
\nonumber \\
&&\qquad + \
{\langle w(\ell_3) PP\rangle_{0}\langle P w(\ell_1)\rangle_{0}
\langle P w(\ell_2)\rangle_{0}\over (\langle PP \rangle_{0})^2 }
\nonumber \\
&&\qquad - \
{\langle w(\ell_1) P\rangle_{0}\langle w(\ell_2) P\rangle_{0}
\langle w(\ell_3) P\rangle_{0}\langle PPP \rangle_{0}
\over (\langle PP \rangle_{0})^3 }\Biggr\}\Bigg\vert_{t=t(\bar t\,)}\ ,
\end{eqnarray}
and so on.
In the above expressions subscript $0$ denotes the genus of the surface
and $P$ is the puncture operator.
Using the known results for the loop correlators when
$g=0$ \cite{oldCF,MSS} one easily checks that our results~(\ref{m1l}),
(\ref{m2l}) and (\ref{m3l}) agree with the direct calculations
of loop correlators in refs.~\cite{ABC,Ko}.

For example, for the $k=2$ model on has explicitly:
\begin{eqnarray} \label{explicit}
\langle\bar w(\ell)\rangle &=&
{1\over\ell^{5/2}}\,\left(1+\ell\bar t^{\,1/3}\right)
\e^{-\ell\bar t^{\,1/3}}\ ,
\nonumber \\
\langle\bar w(\ell_1) \bar w(\ell_2)\rangle &=&
{1\over 2}\,{1\over\sqrt{\ell_1\ell_2}}\,
\left({\ell_1\ell_2\over \ell_1+\ell_2}+\bar t^{\,-1/3}\right)\,
\e^{-(\ell_1+\ell_2)\bar t^{\,1/3}}\ ,
\nonumber \\
\langle\bar w(\ell_1) \bar w(\ell_2) \bar w(\ell_3)\rangle &=&
{1\over 4}\,{1\over\sqrt{\ell_1\ell_2\ell_3}}\,
\Bigl(\ell_1\ell_2\ell_3 \bar t^{\,-1/3} +
(\ell_1\ell_2+\ell_1\ell_3+\ell_2\ell_3)\bar t^{\,-2/3}
\nonumber \\
&+& (\ell_1+\ell_2+\ell_3)\bar t^{\,-1}+ \bar t^{\,-4/3}\Bigr)\
\e^{-(\ell_1+\ell_2+\ell_3)\bar t^{\,1/3}}\ .
\end{eqnarray}
Expanding $\langle\bar w(\ell)\rangle$ in powers of $\ell$,
one finds that the power $\sqrt{\ell}$ is missing, which is related to
the absence of the dimension $\bar\gamma$ operator. Similarly, in the
expansion of $\langle\bar w(\ell_1) \bar w(\ell_2)\rangle$ there are
no terms with $\sqrt{\ell}_1$ or $\sqrt{\ell}_2$, etc.
Thus, instead of (\ref{wexp}) we may write
\begin{equation} \label{mwexp}
\bar w(\ell)={1\over\sqrt{\ell}}\,\bar\sigma_0 +
\sum_{n=1}^{\infty}\ {\ell^{n+1/2} \over \Gamma(n+3/2)}\ \bar\sigma_{n}\ .
\end{equation}
This reflects the fact that the new puncture operator comes from
analytic terms in the conventional model, while the
conventional puncture operator becomes analytic.

Note that the relation~(\ref{Legendre}) between loop correlators in modified
and conventional matrix models is the same as the relation between
the generating functions of connected and one-particle irreducible
Green's functions. This analogy suggests a simple diagrammatic relation
between the two sets of correlators.

Let us now extend the above analysis to surfaces of arbitrary genus.
The saddle-point expansion of eq.~(\ref{double}) generates the
complete genus expansion. It is, however, more convenient to formulate
a set of graphical rules (``Feynman rules'')
\begin{figure}[thb]
\vskip 5pt
\centerline{\epsfxsize 4.0in\epsfbox{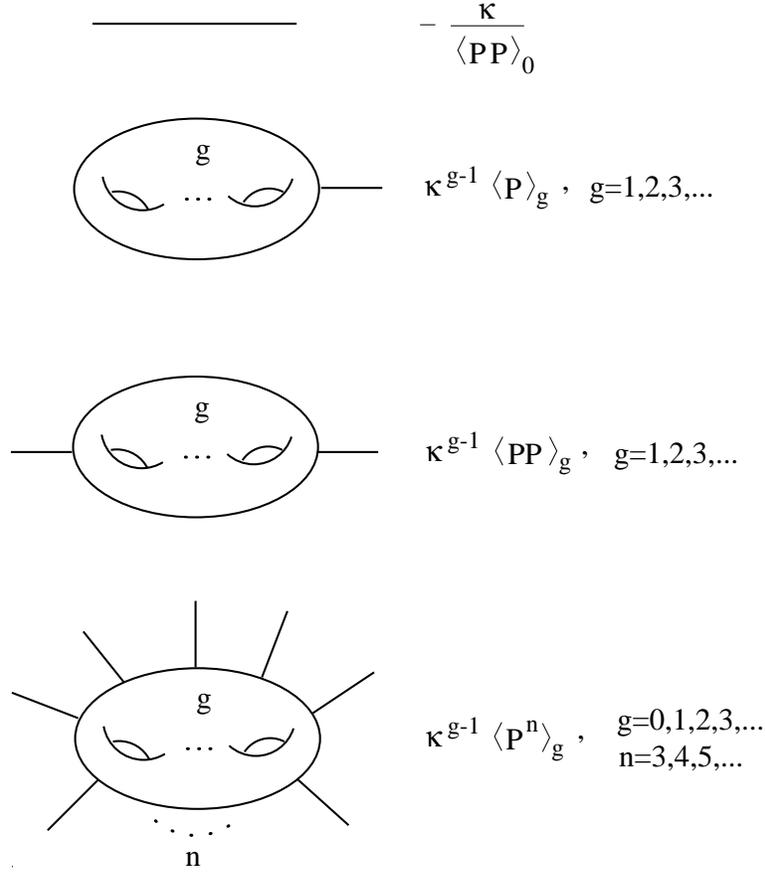}}
\vskip 5pt
\caption{Graphical rules for constructing correlation functions in
the modified matrix models in terms of those in the conventional
models.}
\label{fig:rules}
\end{figure}
which allow us to relate
correlation functions in two models in a simple and geometrically
transparent way. Expanding the exponent in the integrand of
eq.~(\ref{double}) around the saddle point, $t=t_s$, we have,
\begin{equation} \label{exp}
\e^{\bar F}=\int\d t\, {\rm exp}\,\left\{
{t_s\bar t\over\kappa}+{\bar t\over\kappa}\Delta t+ F(t_s) +
{1\over\kappa}\langle P \rangle \Delta t +
\sum_{n\ge 2}{1\over n!}\,{1\over\kappa}\,\langle P^n \rangle
(\Delta t)^n \right\} ,
\end{equation}
where
$$\Delta t\equiv t-t_s\,,\qquad
{1\over\kappa}\,\langle P^n \rangle\equiv F^{(n)}(t_s)\ ,$$
and we have exhibited explicit dependence on the string coupling
constant, $\kappa$.
The saddle point is determined by
\begin{equation} \label{sp}
\bar t+\langle P \rangle =0\ ,
\end{equation}
where $\langle P \rangle$ is the {\em exact} one-point correlation
function of the puncture operator. Since we are eventually interested
in the expansion of $\e^{\bar F}$ in powers of $\kappa$, it is most
convenient to solve eq.~(\ref{sp}) on the sphere,
\begin{equation} \label{sps}
\bar t+\langle P \rangle_0 =0\ ,
\end{equation}
and introduce explicit tadpole terms for surfaces of genus $g\ge 1$.
Similarly, the propagator which we read off the quadratic part in
eq.~(\ref{exp}), $-\kappa/\langle PP \rangle$, depends on
$\kappa$ in a complicated way,
\begin{eqnarray}
-{\kappa\over\langle PP \rangle}&=&
-{\kappa\over\sum_{g\ge 0}\kappa^g\langle PP \rangle_{g}}=
-{\kappa\over\langle PP \rangle_{0}}\Biggl\{ 1-
\sum_{g\ge 1}\kappa^g{\langle PP\rangle_{g}\over\langle PP \rangle_{0}}
\nonumber \\
&+& \sum_{g_1,g_2\ge 1}\kappa^{g_1+g_2}
{\langle PP\rangle_{g_1}\langle PP\rangle_{g_2}\over
(\langle PP \rangle_{0})^2} +\dots \Biggr\}\ .
\end{eqnarray}
Therefore, we choose as a propagator $-\kappa/\langle PP \rangle_0$,
and take higher-genus mass insertion terms as vertices. The complete
list of graphical rules is shown in fig.~\ref{fig:rules}.

\begin{figure}[htbp]
\vskip 10pt
\centerline{\epsfxsize 4.2in\epsfbox{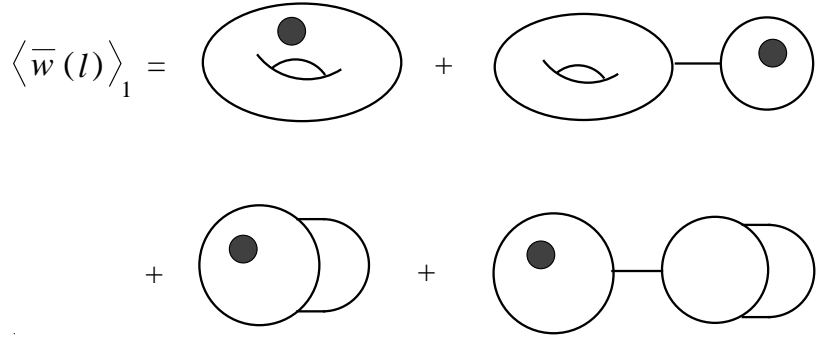}}
\vskip 5pt
\caption{One-loop correlator on the torus.}
\label{fig:1loop1}
\end{figure}

As an example, we show the one-loop correlator on the torus ($g=1$)
in fig.~\ref{fig:1loop1}.
The corresponding analytical expression reads:
\begin{equation}
\langle\bar w(\ell) \rangle_{1} =
\langle w(\ell) \rangle_{1} -
{\langle P\rangle_{1}\langle w(\ell)P\rangle_{0}\over\langle PP\rangle_{0}}
-{1\over 2}\,{\langle w(\ell)PP\rangle_{0}\over\langle PP\rangle_{0}}
+ {1\over 2}\,{\langle w(\ell)P\rangle_{0}
\langle PPP \rangle_{0}\over (\langle PP\rangle_{0})^2}\ .
\end{equation}

The reduction formulae do not work for correlation functions involving
puncture operator. Instead, one uses the chain rule which follows from
eq.~(\ref{sps}):
\begin{equation}
\langle\bar{P}^{n}\cdots\rangle =
{\partial^{n}\over\partial\bar{t}^{n}}\, \langle\cdots\rangle
=\left({\partial t_s\over \partial\bar{t}}\,
{\partial\over\partial t_s}\right)^{n}
= \left(-{1\over\langle PP\rangle_{0}}\,
{\partial\over\partial t_s}\right)^{n} \ \langle\cdots\rangle \ ,
\end{equation}
with particular cases
\begin{equation}
\langle\bar{P}\rangle_0 =
\kappa {\partial\over\partial\bar{t}}\,\bar{F}_0=
{\partial\over\partial\bar{t}}\,\Bigl(\bar{t}t_s+F_0(t_s)\Bigr)
= t_s=\bar{t}^{\,1-\bar\gamma} \ ,
\end{equation}
and
\begin{equation}
\langle\bar{P}\bar{P}\rangle_0 =-{1\over\langle PP\rangle_0}\ .
\end{equation}
A higher genus example is the one-punctured torus:
\begin{equation}
\langle\bar{P}\rangle_1 = {\partial\over\partial\bar{t}}\,\bar{F}_1 =
-{\langle P\rangle_1 \over \langle PP \rangle_0} +
{1\over 2}\, {\langle PPP \rangle_0\over \langle PP \rangle_0}\ .
\end{equation}

\section{Virasoro Constraints and Recursion Relations}

One of the most important mathematical properties of the $c<1$ matrix
models is the existence of Virasoro constraints on the partition
function (or appropriate generalizations for multi-matrix models).
These constraints are equivalent to the loop equations and are related
to the integrability property of these models, both on the lattice and
in the continuum. When expressed in terms of correlators of local
scaling operators, they take the form of recursion relations identical
to those defining topological two-dimensional gravity. A nice
geometrical picture arises at the multicritical points with $\gamma_k
= -1/k$: all correlators of scaling operators $\sigma_m$ with $m>k-2$
reduce to contact terms at the boundaries of moduli space and  can
be solved in terms of the correlators of $\sigma_m$ with $m<k-1$. This
defines the notion of gravitational primaries and descendants, a
gravitational version of relevant and irrelevant perturbations of the
critical point.

The Virasoro constraints in the continuum take the form~\cite{DVV,FKN}
(defining $t_0\equiv t$):
\begin{equation} \label{virc}
L_n Z(t_0, \{t\}) = 0 \ , \quad n\ge -1\ ,
\end{equation}
where $\{t\} = t_1, t_2, \dots$,  and
$Z = \e^F$ is the disconnected sum over continuous surfaces.\footnote{The
continuum limit of matrix models with even potentials
induces a doubling of the degrees of freedom, so that
the Virasoro constraints act on the square root of the matrix model
partition function in that case.} The differential
operators $L_n$ are given by
\begin{eqnarray} \label{virs}
L_{-1} &=& \sum_{m\ge 1} \left(m+{1\over 2}\right) t_m {\partial \over
\partial t_{m-1}} + {t_0^2 \over 8\kappa} \ ,
\nonumber \\
L_0 &=& \sum_{m\ge 0} \left(m+{1\over 2}\right) t_m
{\partial \over \partial t_m} + {1\over 16}  \ ,
\nonumber \\
L_n &=& \sum_{m\ge 0} \left(m+{1\over 2}\right) t_m {\partial \over
\partial t_{m+n}} + {\kappa \over 2} \sum_{m=1}^{n} {\partial^2 \over \partial
t_{m-1} \partial t_{n-m}}\ ,
\end{eqnarray}
where we made explicit the dependence on $\kappa$, the string loop
expansion parameter. These operators satisfy a centerless Virasoro
algebra
\begin{equation} \label{viral}
[L_n, L_m ] = (n-m) L_{n+m} \ .
\end{equation}
In the modified matrix models, the partition function (and in general
any disconnected correlator) is defined in terms of the corresponding
object in the standard matrix model by the Laplace transform
\begin{equation} \label{otra}
{\bar Z}({\bar t_0}, \{t\}) = \int \d t_0 \, \e^{t_0 {\bar t_0}/\kappa}
Z(t_0, \{t\})  \equiv {\cal L} [Z] \ .
\end{equation}
This formula makes sense as the saddle point or genus expansion and,
as was discussed in the previous section, any correlation function of loops
or scaling operators in  models of ``touching" surfaces
may be decomposed into sums of products of correlators
of conventional models. It is then clear that the conventional
recursion relations imply certain modified recursion relations
in the new models, perhaps with a similar topological interpretation.
In fact, such relations follow immediately from (\ref{otra})
if we define a set of modified Virasoro operators
by the operator identity
\begin{equation} \label{com}
\bar L_n {\cal L} = {\cal L} L_n\ ,
\end{equation}
thus satisfying
\begin{equation} \label{barvir}
{\bar L}_n \, {\bar Z}({\bar t}_0, \{t\}) = 0 \ , \quad n\ge -1\ .
\end{equation}
The new operators are related to those in (\ref{virs})  by the
transformations
\begin{equation} \label{trans}
t_0 \rightarrow \kappa {\partial \over \partial {\bar t}_0} \ , \qquad
{\partial \over \partial t_0} \rightarrow -{\bar t_0 \over \kappa}\ .
\end{equation}

One can explicitly check that the $\bar L_n$ operators so defined
satisfy the same Virasoro algebra as $L_n$. A simple proof  of this
fact follows from the abstract Fock representation introduced in
ref.~\cite{DVV}. The Virasoro operators (\ref{virs})
can be interpreted in terms of the
energy-momentum tensor of a twisted scalar field on the circle, with
mode expansion
\begin{equation}
\partial \phi(z) = \sum_{n} \alpha_{n+{1\over 2}} z^{-n-3/2}\ ,
\end{equation}
under the following substitutions for $n\ge 0$:
\begin{equation}
\alpha_{-n-{1\over 2}}={1\over\sqrt{\kappa}}\left(n+{1\over 2}\right)
t_n\ , \qquad \alpha_{n+{1\over 2}}=\sqrt{\kappa}
{\partial\over\partial t_n}\ .
\end{equation}
It is easy to see that the Fock space representantion of the modified
Virasoro operators is linearly related to the previous one, $\bar
\alpha_{n+1/2} = \alpha_{n+1/2}$ for $n\neq 0,1$ and $\bar
\alpha_{1/2} = -2 \alpha_{-1/2}$, $\bar \alpha_{-1/2} = {1\over 2}
\alpha_{1/2}$. In fact, it is a Bogoliubov transformation, since it
conserves the canonical algebra
\begin{equation}
[\alpha_{n+{1\over 2}}, \alpha_{m-{1\over 2}}] =
[\bar \alpha_{n+{1\over 2}} , \bar \alpha_{m-{1\over 2}} ]=
\left(n+{1\over 2}\right) \delta_{n+m,0}
\end{equation}
from which one derives  the Virasoro algebra (\ref{viral}) for the $\bar L_n$
operators.

Recursion relations are easily obtained starting from the general
identity:
\begin{equation} \label{identity}
\prod_{j\in S} {\partial\over \partial t_j} \left( \e^{-\bar F} \, \bar
L_n \, \e^{\bar F}\right) = 0 \ ,
\end{equation}
expanded in powers of the string coupling at the $k$th multicritical
point, $t_n = -\delta_{n,k} / (2k+1)$, $n\ge 1$, with string
susceptibility $\bar \gamma_k = 1/(k+1)$. Let us consider for simplicity
the case in which there is no puncture operator $\bar\sigma_0 = \bar P$ in
the set $S$, and use the notation $\bar\sigma_S \equiv \prod_{S}
\bar\sigma_j$. Neglecting some analytic terms in the couplings, the
$L_{-1}$ or puncture equation takes the form:
\begin{equation}
\langle \bar\sigma_{k-1} \bar\sigma_S\rangle_g =\sum_{j\in S}(2j+1)\langle
\bar\sigma_{j-1} \bar\sigma_{S-j}\rangle_g
+ {1\over 4} \langle \bar P \bar P \bar\sigma_S \rangle_{g-1} + {1\over 8}
\sum_{X\cup Y} \sum_{g_1 +g_2 =g} \langle \bar\sigma_X \bar P\rangle_{g_1}
\langle \bar P \bar\sigma_Y \rangle_{g_2} \ .
\end{equation}
The first term on the r.h.s. represents operator contact terms and is
identical to the conventional counterpart. However, in the modified model we
have additional factorization terms from the contribution of the
puncture operator at the boundaries of moduli space, where a genus $g$
surface degenerates into two $g_1 + g_2 = g$ surfaces or a $g-1$
surface by pinching a handle.

The $L_0$ or dilaton equation is identical to the conventional one, up to a
sign flip of the bulk term, proportional to $\bar t_0$:
\begin{equation}
\langle \bar\sigma_k \bar\sigma_S \rangle_g = \sum_{j\in S} (2j+1)\langle
\bar\sigma_j \bar\sigma_{S-j} \rangle_g - \bar t_0 \langle
\bar P \bar\sigma_S \rangle_g \ .
\end{equation}
It is interesting that this equation is insensitive to the puncture
factorization. The higher $n\ge 1$ equations take the following
form:
\begin{eqnarray}
\langle \bar\sigma_{k+n} \bar\sigma_S\rangle_g &=&\sum_{j\in S}(2j+1)\langle
\bar\sigma_{j+n} \bar\sigma_{S-j}\rangle_g -2\bar t_0 (1-\delta_{n,1})\langle
\bar\sigma_{n-1} \bar\sigma_S \rangle_g
\nonumber \\
&+& \langle \bar P \bar\sigma_n \bar\sigma_S\rangle_{g-1}+\sum_{m=2}^{n-1}
\langle \bar\sigma_{m-1} \bar\sigma_{n-m} \bar\sigma_S \rangle_{g-1}
\nonumber \\
&+& {1\over 2} \sum_{X\cup Y} \sum_{g_1 +g_2 =g} \left(
\langle \bar\sigma_X \bar P\rangle_{g_1}\langle\bar\sigma_n\bar\sigma_Y
\rangle_{g_2}+\sum_{m=2}^{n-1}\langle
\bar\sigma_X \bar\sigma_{m-1} \rangle_{g_1} \langle \bar\sigma_{n-m}
\bar\sigma_Y \rangle_{g_2} \right) .
\end{eqnarray}
Again, the operator contact term as well as the factorization terms not
involving the puncture operator are identical to the conventional case.
The factorization for the puncture is different though: it comes with
an extra factor of $1/2$, and the conjugated operator
at the degeneration neck is $\bar\sigma_n$ instead of $\bar\sigma_{n-1}$. The
bulk term is also slightly different.

These equations are still recursion relations for the gravitational
descendants, in the usual sense, and
can be derived from the corresponding relations in the conventional
models using the
diagrammatic method explained in the previous section. This represents
a non-trivial check of both the recursion relations and the
diagrammatic rules for touching surfaces, due to the complicated
pattern of cancellations involved. For this reason it is perhaps
surprising that the final result is so similar to the conventional
case, with only
some modifications in the way the new puncture operator enters in
factorization diagrams and bulk terms. We interpret this fact as
evidence for an underlying topological field theory description,
involving only slight changes from the usual one, perhaps similar to
the change of dressing branches, advocated for the Liouville model.
Presumably, the r{\^o}le of the pure topological point, $k=1$, is played
here by the pure polymer phase with $\bar \gamma_1 = 1/2$.

To conclude this section, we remark that one can also write
integrated forms of the $L_{-1}$ and $L_0$ equations acting on loops,
measuring boundary lengths and overall dilatations. To this end, we
simply insert the loop operator
\begin{equation}
\bar w(\ell) = {1 \over \sqrt{\ell}} {\partial \over \partial \bar
t_0} + \sum_{n\ge 1} {\ell^{n+1/2} \over \Gamma(n+3/2)} {\partial \over
\partial t_n}
\end{equation}
on the left hand side of eq.~(\ref{identity}).

We find some differences with respect to the conventional models.
The dilaton operator, which at the $k$th critical point is
$\bar\sigma_k$, satisfies
\begin{equation}
{1\over 2}\langle \bar\sigma_k\bar w(\ell)\rangle=\left( -{1\over 2} \bar t_0
{\partial \over \partial \bar t_0} + \ell {\partial \over \partial
\ell} \right) \langle \bar w(\ell)\rangle \ .
\end{equation}
Comparing with the conventional models, the sign of the bulk term is
flipped. In the modified models we find additional
non-local terms  associated
with the new puncture contributions to the $L_{-1}$ constraint,
as well as an inhomogeneus term, proportional to the one-puncture
function,
\begin{equation}
{1\over 2}\langle \bar\sigma_{k-1} \bar w(\ell)\rangle =
\ell \langle \bar w(\ell)\rangle
- \sqrt{\ell} \langle \bar P\rangle +{1\over 8} \left( \kappa \langle \bar
P\bar P \bar w(\ell)\rangle + 2 \langle \bar P\rangle\langle \bar P \bar
w(\ell)\rangle\right)\ ,
\end{equation}
where we have neglected some analytic terms.
Thus, it seems that a local
boundary operator is absent in the modified models.
($\sigma_{k-1}$  acts as a boundary operator in the conventional models
due to the relation
${1\over 2}\langle\sigma_{k-1} w(\ell)\rangle=\ell\langle w(\ell)\rangle$.)
There are interesting generalizations of the above exercise.
For example, one may consider $W$-constraints for
modified multi-matrix models. Also, other operators can be tuned as
wormhole sources. In the most general case one has a multiple Laplace
transform and still a Bogoliubov transformation in the Fock space
representation.

\section{Changing Gravitational Dimensions}

In the previous sections we studied modified one-matrix models with the
simplest type of trace squared term, $(\Tr\Phi^4)^2$.
We showed that the effect of this term is to change the branch of
gravitational dressing of the lowest dimension operator in the
theory, which corresponds to coupling constant $t$.
In this section we check that more complicated trace-squared terms
alter the gravitational dimensions of other scaling
operators. In particular, we check in detail that formula (\ref{newdouble}),
first derived in ref. \cite{KlHa}, applies to gravitational
descendants.

Although our approach is general, we focus for simplicity on the $k=2$
one-matrix model, whose matrix potential is
\begin{equation}
S_0 (\Phi) = N\Tr ({1\over 2}\Phi^2 -\lambda\Phi^4) \ .
\end{equation}
The continuum limit is achieved as $\Delta=\lambda_c-\lambda\rightarrow
0$.
It is convenient to introduce the first scaling operator of the form
\begin{equation}
\sigma_1= N\Tr\left(\Phi^2 -{1\over 12}\Phi^4\right )-
N^2 (1-16\Delta-2304\Delta^2)\ .
\end{equation}
This operator has gravitational dimension $d=1/2$, and its
connected genus zero correlation functions are given by
\begin{eqnarray} \label{mir}
&&
\langle\sigma_1\rangle= N^2 \CO(\Delta^{5/2})\ ,
\nonumber \\
&&
\langle\sigma_1 \sigma_1\rangle= N^2 ({4\over 3}-1024\sqrt 3\Delta^{3/2}
 +\CO(\Delta^2))\ ,
\nonumber \\
&&
\langle\sigma_1 \sigma_1\sigma_1\rangle= N^2 (1536\Delta
 +\CO(\Delta^{3/2}))\ ,
\nonumber \\
&&
\langle\sigma_1 \sigma_1\sigma_1\sigma_1\rangle= N^2 (-768\sqrt 3\Delta^{1/2}
 +\CO(\Delta))\ ,\ {\rm etc.}
\end{eqnarray}
The purpose of the $\Phi$-independent term in $\sigma_1$ is to
remove a non-universal analytic term from the one-point function.
Apart from this it has no effect.

Now we consider a modified model with the action
\begin{equation}
S_0 (\Phi) -\tau_1 \sigma_1 -{g\over 2N^2} (\sigma_1)^2 \ ,
\end{equation}
where we have introduced coupling constant $\tau_1$
in order to study correlation functions of $\sigma_1$.
Its partition function may be written as
\begin{equation}
Z\sim
\int_{-\infty}^\infty \d v\, \e^{-{N^2 v^2\over 2g}}
\int {\cal D} \Phi\, \e^{- S_0+(v+\tau_1)\sigma_1}\ .
\end{equation}
Defining a shifted variable $u=v+\tau_1$, we perform the matrix
integral first and reduce the modified free energy to
\begin{eqnarray}
&&
\bar F=  F_0(t) + \log \int \d u\, \e^{f(u)} \ ,\nonumber \\
&& f(u)= -{N^2\over 2 g} (u^2- 2u \tau_1 + \tau_1^2)+
{4\over 3} N^2 u^2 +
F_1 (t, t_1)\ ,
\end{eqnarray}
where the scaling variables $t$ and $t_1$ are defined through
\begin{equation}
t\sim \Delta N^{2/(2-\gamma)}\ ,\qquad\qquad
t_1= u N^{2(1-d)/(2-\gamma)} \ .
\end{equation}
In this specific case
$\gamma=-1/2$ and $d=1/2$. Of crucial importance is the fact
the $F_1$ depends only on the scaling variables and has the
form
\begin{equation}\label{sf}
F_1 (t, t_1)=\sum_{n=2}^\infty a_n t_1^n t^{(5-n)/ 2} \ ,
\end{equation}
where $a_n$ are constants (the first three of them may be
read off eq. (\ref{mir})).
This follows from miraculous vanishing
of certain potentially harmful non-universal terms in eq.
(\ref{mir}). For example, had the three-point function of $\sigma_1$
started with a non-universal term of order $\Delta^0$,
eq. (\ref{sf}) would not be valid!

If we now set $g=3/8$ and introduce
the scaling variable
\begin{equation}
\bar t_1= {8\over 3}\tau_1 N^{2(1-\bar d)/(2-\gamma)} \ ,
\end{equation}
we arrive at the following expression for
the universal part of the modified free energy,
\begin{equation}
\bar F (t,\bar t_1) =\log \int_{-\infty}^\infty \d t_1\,
\e^{t_1 \bar t_1 + F(t, t_1)} \ ,
\end{equation}
Here $F(t, t_1)$ is the universal part of the conventional sum over
surfaces. Thus, we find that the gravitational dimension of $\sigma_1$
has changed from $d=1/2$ to
$\bar d= \gamma-d=-1$. This provides a counterexample to the claim
of ref.~\cite{Sh} that eq. (\ref{newdouble}) applies only to operators with
$d\le (1+\gamma)/3$.

While in ref. \cite{KlHa} eq. (\ref{newdouble}) was derived
for gravitational primary fields, we have just checked it for
a gravitational descendant. Our example shows that these operators
have to be defined in such a way that their correlation functions
do not contain certain non-universal contributions. We believe that
this is in fact always possible. As further evidence, we present
results for the dilaton operator,
\begin{equation}
\sigma_2= N\Tr \left (2\Phi^2 -{1\over 3}\Phi^4+
{1\over 60}\Phi^6\right )-N^2\left({8\over 5}-{64\over 5}\Delta\right) .
\end{equation}
This operator has gravitational dimension $d=1$, and its
connected genus zero correlation functions are given by
\begin{eqnarray} \label{mird}
&&
\langle\sigma_2\rangle\sim N^2 \Delta^{5/2}\ ,
\nonumber \\
&&
\langle\sigma_2 \sigma_2\rangle= N^2 ({32\over 15} +\CO(\Delta^{5/2}))\ ,
\nonumber \\
&&
\langle\sigma_2 \sigma_2\sigma_2\rangle \sim N^2 \Delta^{5/2}\ ,
\nonumber \\
&&
\langle\sigma_2 \sigma_2\sigma_2\sigma_2\rangle
\sim N^2 \Delta^{5/2}\ ,\ {\rm etc.}
\end{eqnarray}
Once again, the unwanted non-universal terms vanish!
Repeating the steps carried out for $\sigma_1$, we may now establish
the validity of (\ref{newdouble}) for $\sigma_2$.
This relation, and the change of gravitational dimension it implies,
appear to be completely general.

\section{Correlators of Dressed Primary Fields}

In the previous section we studied correlation functions of
gravitational descendants in modified one-matrix models.
In this section we turn to gravitationally dressed primary fields.
Although our approach is general, we mostly discuss some simple
special cases: genus zero two- and three-point functions of the order parameter
fields in unitary minimal models coupled to gravity, as well as
the four-point function for $c=1$. We find that the correlators in modified
matrix models agree with those of negatively dressed operators in
Liouville theory, provided that the latter are obtained with the
simplest plausible analytic continuation prescription.
Although a far better understanding of the
Liouville theory calculations is desirable, we feel that our findings
suggest a general pattern for interpreting the
modified matrix models in terms of the negatively dressed operators.

The starting point for our calculations
is the relation (\ref{newdouble}) between the genus zero
free energy $\bar F$ of the modified matrix model and that
of the original matrix model, $F$:
\begin{equation} \label{lap}
\bar F(\bar t_1,\bar t_2,...) = \sum_it_i\bar t_i + F(t_1,t_2,...) ,
\qquad\qquad {\partial F\over\partial t_j} = -\bar t_j\ ,
\end{equation}
where $t_i$ and $\bar t_i$ represent the coupling constants by which the
model is perturbed (it is implicit that $F$ and $\bar F$ also depend
on the basic coupling constant $t$, which appears in the
Liouville action). Since $t_i$ correspond to
dressed primary fields, $F$ has the expansion
\begin{equation} \label{sarah}
F(t_i) = {1\over2}\Delta_i t_i^2 + {1\over6} c_{ijk} t^i t^j t^k
+{1\over24}d_{ijkl} t^i t^j t^k t^l + \dots
\end{equation}
where $\Delta_i, c_{ijk}, d_{ijkl}$ are the ordinary two-, three- and
four-point functions.
$\bar F$ in (\ref{lap}) can be evaluated order by order in the $\bar t_i$
by expanding $t_i\bar t_i+F(t_i)$ around the saddle point.

Let us begin by discussing the two- and three-point functions.
First assume that there is only one parameter $t_i$.
Then the Legendre transform fixes $t_i$ at
\begin{equation}
t_{i0}= -{1\over\Delta_i}\bar t_i-{c_{iii} \over 2\Delta_i^3}
\bar t_i^{\,2} + \dots
\end{equation}
The genus zero part of $\bar F$ is just the value of
$t_i\bar t_i+F(t_i)$ at $t_{i0}$. One finds in this case
\begin{equation}
\bar F(\bar t_i) = {1\over2}\Delta_{\bar\imath} \bar t_i^{\,2} + {1\over6}
c_{\bar\imath\bar\imath\bar\imath}\bar t_i^{\,3} + \dots
\end{equation}
with modified two- and three-point functions
\begin{equation} \label{tim}
\Delta_{\bar\imath} = -{1\over\Delta_i}\ , \qquad
c_{\bar\imath\bar\imath\bar\imath} =
-{c_{iii} \over \Delta_i^3}\ .
\end{equation}
Next, assume that there are two parameters $t_i,t_k$ and one wants to
modify one of them, i.e., one wants to find
$\bar F(\bar t_i,t_k)$ to cubic order. Now the saddle point is at
\begin{equation}
t_{i0}= \left(-{1\over\Delta_i}+{c_{iik} \over 3\Delta_i^2}t_k\right)
\bar t_i + \dots
\end{equation}
In this case,
\begin{equation}
\bar F(\bar t_i,t_k) = {1\over2}\Delta_{\bar\imath}\bar t_i^{\,2}
+ {1\over2} \Delta_k t_k^2 + {1\over6} c_{\bar\imath\bar\imath k}
\bar t^i \bar t^i t^k + {1\over6} c_{\bar\imath kk}\bar t^i t^k t^k + \dots
\end{equation}
with
\begin{equation} \label{tom}
\Delta_{\bar\imath} = -{1\over\Delta_i}\ , \qquad
   c_{\bar\imath\bar\imath k} = +{c_{iik} \over \Delta_i^2}\ , \qquad
   c_{\bar\imath kk} = -{c_{ikk} \over \Delta_i}\ .
\end{equation}
The generalization is straightforward: each time we Legendre transform
from $t_i$ to $\bar t_i$ the two- and three-point functions change
according to
\begin{equation} \label{theo}
\Delta_{i} \rightarrow
\Delta_{\bar\imath} = -{1\over\Delta_i}\ , \qquad
   c_{ijk} \rightarrow
   c_{\bar\imath j k} = -{c_{ijk} \over \Delta_i}\ .
\end{equation}
The absolute values of the correlators are normalization dependent, so it
is useful to define the normalization independent quantity
as in ref. \cite{GoLi}:
\begin{equation} \label{maja}
X_{ijk} = {c_{ijk}^2 \over \Delta_i\Delta_j\Delta_k} F\ ,
\end{equation}
where $F$ is the free energy evaluated at $t_i=0$.
One easily sees from eq.~(\ref{theo})
that $X$ has a simple property: it just switches sign each time one
of the  external operators is modified, i.e.
\begin{equation} \label{signc}
X_{\bar\imath  jk} =-X_{ijk}\ ,\qquad
X_{\bar\imath\bar\jmath k} = X_{ijk}\ ,\qquad
X_{\bar\imath\bar\jmath\bar k} =-X_{ijk}\ .
\end{equation}

Let us compare this behavior with what one would expect from Liouville
theory, if the modified matrix model operators $\bar{\cal O}_i$ were
identified with the negatively dressed operators.
Consider the $q$th minimal model with central charge
$$c=1- {6\over q(q+1)}\ ,$$
coupled to gravity. The gravitationally
dressed operators on the diagonal of the Kac table are
\begin{equation}
{\cal O}_i=\psi_{r_i,r_i}\, \e^{\beta_i^+\phi}\ ,
\qquad\qquad \bar{\cal O}_i= \psi_{r_i,r_i}\, \e^{\beta_i^-\phi}\ ,
\end{equation}
with
$$
\beta_i^\pm = -{Q\over2}\pm\omega_i\ ,\qquad
\omega_i={r_i\over{\sqrt{2q(q+1)}}}\ .
$$
We note
that switching from the $\beta_i^+$ to the $\beta_i^-$ dressing corresponds
to continuing $r_i\rightarrow -r_i$.
The dressed unity $e^{\alpha_+\phi}$
corresponds to the puncture operator with $r_i=1$.
As shown in section 2,
if we Legendre transform with respect to the cosmological constant
$t$, as in eq. (\ref{Legendre}), we find a new theory with
Liouville potential $e^{\alpha_-\phi}$.

First, note that the scaling behavior of the modified correlators
(\ref{tim}) and (\ref{tom}) is that of
operators with the $\beta^-_i$ dressing. This has already
been noted in ref. \cite{KlHa} and in section 4.
Next, let us compare the coefficients.
Consider the quantities $X_{ijk}$, defined in (\ref{maja}).
They can be derived from the fact that the three-point function
of the properly normalized operators can be
written as \cite{3pt}
\begin{equation} \label{willy}
\langle {\cal O}_i {\cal O}_j {\cal O}_k \rangle=
t^{-{Q\over\alpha_+}-{\beta_i^++\beta_j^++\beta_k^+
\over \alpha_+}}\ .
\end{equation}
Since insertions of the puncture operator $P$ are produced by
$-\partial_t$, we can infer from (\ref{willy}) and from the scaling
behavior the two-point function and the partition function:
\begin{equation}
-\partial_t \langle {\cal O}_k {\cal O}_k \rangle =
\langle P {\cal O}_k {\cal O}_k \rangle=
t^{-{Q\over\alpha_+}-2{\beta_k^+ \over \alpha_+}-1}\ \ \Rightarrow \ \
\langle {\cal O}_k {\cal O}_k \rangle =
{\alpha_+\over Q+2\beta_k^+}\,t^{-{Q\over\alpha_+}-
2{\beta_k^+ \over\alpha_+}}\ ,
\end{equation}
and
\begin{equation}
\partial_t^2 F= \langle PP\rangle= {\alpha_+ \over Q+2\alpha_+}
t^{-{Q\over\alpha_+}-2} \ \ \Rightarrow \ \
F={1\over {Q\over\alpha_+}({Q\over\alpha_+}+1)({Q\over\alpha_+}+2)}\,
t^{-{Q\over\alpha_+}}\ .
\end{equation}
We thus obtain
\begin{equation} \label{invx}
X_{ijk}={\langle {\cal O}_i {\cal O}_j {\cal O}_k \rangle^2  F\over
\langle {\cal O}_i {\cal O}_i\rangle \langle
{\cal O}_j {\cal O}_j\rangle \langle {\cal O}_k {\cal O}_k \rangle}
={(Q+2\beta_i^+) (Q+2\beta_j^+) (Q+2\beta_k^+)\over
Q (Q+\alpha_+) (Q+2\alpha_+) }
\sim r_i r_j r_k\ .
\end{equation}
This is the result of Goulian and Li \cite{GoLi}.
In order to calculate $X_{\bar\imath jk}$ we assume that it is given
by the above equation with
$\beta^+$ replaced by $\beta^-$. This procedure is consistent with
all the tricks used in the Liouville theory calculations \cite{GoLi}.
The result is remarkably simple: $X_{\bar\imath jk}=- X_{ijk}$,
in agreement with (\ref{signc}). In general,
$X_{ijk}$ simply switches sign each time one of the Liouville
dressings is modified from $\beta^+$ to $\beta^-$.
It appears that we have found an explanation for the changes of sign
caused by Legendre transform (\ref{lap}). Remarkably, they have the
same effect on correlation functions as changes in the sign of
the Liouville energies.

Equally remarkable is the effect on (\ref{invx}) of Legendre
transforming with respect to the cosmological constant $t$ only, as in
eq.~(\ref{Legendre}). In this modified matrix model we find
\begin{equation}
\bar F={1\over {Q\over\alpha_+}({Q\over\alpha_+}+2)}\,
(C\bar t\,)^{-{Q\over\alpha_-}}\ ,
\end{equation}
where $C=({Q\over\alpha_+}+1)({Q\over\alpha_+}+2)$.
Correlation functions of order parameters other than the puncture
become ``one-puncture irreducible'' and may be computed using the same
rules as in eqs.~(\ref{m1l})--(\ref{m3l}). The modifications are trivial
because one-point functions vanish, and we arrive at
\begin{eqnarray}
&&
\langle {\cal O}_k {\cal O}_k \rangle =
{\alpha_+\over Q+2\beta_k^+}\,(C\bar t\,)^{-{Q\over\alpha_-}-
2{\beta_k^+ \over\alpha_-}}\ ,
\nonumber \\
&&
\langle {\cal O}_i {\cal O}_j {\cal O}_k \rangle=
(C\bar t\,)^{-{Q\over\alpha_-}-{\beta_i^++\beta_j^++\beta_k^+
\over \alpha_-}}\ .
\end{eqnarray}
This leads to
\begin{eqnarray}
\bar X_{ijk} &=& {\langle {\cal O}_i {\cal O}_j {\cal O}_k \rangle^2
\bar F\over \langle {\cal O}_i {\cal O}_i\rangle \langle
{\cal O}_j {\cal O}_j\rangle \langle {\cal O}_k {\cal O}_k \rangle}
={(Q+2\beta_i^+) (Q+2\beta_j^+) (Q+2\beta_k^+)\over
Q \alpha_+ (Q+2\alpha_+) } \nonumber \\
&=& {(Q+2\beta_i^+) (Q+2\beta_j^+) (Q+2\beta_k^+)\over
Q (Q+\alpha_-) (Q+2\alpha_-) } \ .
\end{eqnarray}
Thus, the sole effect of the Legendre transform with respect to $t$ is
to replace $\alpha_+$ by $\alpha_-$ in eq.~(\ref{invx})!
This precisely agrees with the idea that the dressing of the Liouville
potential has changed from $e^{\alpha_+\phi}$ to $e^{\alpha_-\phi}$.

To generalize our observations to the four-point function, it is useful to
note that the relation  between $F(t_i)$ and
$\bar F(\bar t_i)$ is the same as that between
$W(j)$, the generating function of the connected diagrams,
and  $S(\phi)$, the effective action,
\begin{equation}
\e^{S(\phi)} = \int \d j\ \e^{j\phi+W(j)}\ .
\end{equation}
At tree level $S(\phi)$ is identical to
the generating function $\Gamma(\phi)$ of one-particle irreducible diagrams.
Based on this analogy, we can graphically represent the relation between
modified correlators (black circles) and original correlators (white circles)
in terms of Feynman diagrams. Let us illustrate this with some examples:

{}\begin{picture}(440,140)(10,15)

\put(75,98){=\ \ --}
\put(158,115){--1}
\put(80,28){=}
\put(440,98){(A)}
\put(440,28){(B)}

\put(0,70) {\begin{picture}(90,60) \put(30,30){\circle*{5}}
\put(15,30){\line(1,0){30}} \put(3,28){$\bar\imath$} \put(50,28){$\bar\imath$}
\end{picture}}

\put(100,70)
 {\begin{picture}(90,60)
\put(30,30){\circle{5}} \put(15,30){\line(1,0){13}}
\put(33,30){\line(1,0){13}} \put(3,28){$(\, i$} \put(50,28){$i\ )$}
\end{picture}}

\put(0,0)
 {\begin{picture}(90,60)
\put(30,30){\circle*{10}} \put(30,30){\line(1,0){30}}
\put(30,30){\line(-1,1){20}} \put(30,30){\line(-1,-1){20}}
\put(3,50){$\bar\imath$} \put(3,3){$\bar\jmath$} \put(62,28){$\bar k$}
\end{picture}}

\put(100,0)
 {\begin{picture}(90,60)
\put(30,30){\circle{10}} \put(35,30){\line(1,0){25}}
\put(26,26){\line(-1,-1){16}} \put(26,34){\line(-1,1){16}}
\put(45,30){\circle*{5}} \put(20,20){\circle*{5}} \put(20,40){\circle*{5}}
\put(3,50){$\bar\imath$} \put(3,3){$\bar\jmath$} \put(62,28){$\bar k$}
\end{picture}}

\end{picture}

\vskip 0.2in

{}\begin{picture}(450,140)(10,15)
\put(0,80)
 {\begin{picture}(60,60)
\put(30,30){\circle{10}} \put(26,26){\line(-1,-1){16}}
\put(26,34){\line(-1,1){16}} \put(34,34){\line(1,1){16}}
\put(34,26){\line(1,-1){16}} \put(3,50){$i$} \put(3,3){$j$}
\put(50,3){$k$} \put(51,50){$l$}
\end{picture}}

\put(100,80)
 {\begin{picture}(60,60)
\put(30,30){\circle{10}} \put(26,26){\line(-1,-1){16}}
\put(26,34){\line(-1,1){16}} \put(34,34){\line(1,1){16}}
\put(34,26){\line(1,-1){16}} \put(3,50){$i$} \put(3,3){$j$}
\put(50,3){$k$} \put(51,50){$l$}
\end{picture}}

\put(200,80)
 {\begin{picture}(90,60)
\put(30,30){\circle{10}} \put(60,30){\circle{10}}
\put(45,30){\circle*{5}} \put(35,30){\line(1,0){20}}
\put(64,34){\line(1,1){16}} \put(64,26){\line(1,-1){16}}
\put(26,26){\line(-1,-1){16}} \put(26,34){\line(-1,1){16}}
\put(3,50){$i$} \put(3,3){$j$} \put(80,3){$k$} \put(81,50){$l$}
\put(40,34){$\bar m$}
\end{picture}}

\put(80,108){$\rightarrow$}
\put(180,108){+}

\put(0,0)
 {\begin{picture}(60,60)
\put(30,30){\circle*{10}} \put(30,30){\line(1,1){20}}
\put(30,30){\line(1,-1){20}} \put(30,30){\line(-1,1){20}}
\put(30,30){\line(-1,-1){20}} \put(3,50){$\bar\imath$}
\put(3,3){$j$} \put(50,3){$k$} \put(51,50){$l$}
\end{picture}}

\put(100,0)
 {\begin{picture}(60,60)
\put(30,30){\circle{10}} \put(20,40){\circle*{5}}
\put(26,26){\line(-1,-1){16}} \put(26,34){\line(-1,1){16}}
\put(34,34){\line(1,1){16}} \put(34,26){\line(1,-1){16}}
\put(3,50){$\bar\imath$} \put(3,3){$j$} \put(50,3){$k$} \put(51,50){$l$}
\end{picture}}

\put(80,28){=}
\put(440,108){(C)}
\put(440,28){(D)}
\end{picture}

{}\begin{picture}(450,80)(10,15)
\put(80,28){=} \put(180,28){+}
\put(300,28){+\ \ \ \  crossing \ terms}

\put(0,0)
 {\begin{picture}(60,60)
\put(30,30){\circle*{10}} \put(30,30){\line(1,1){20}}
\put(30,30){\line(1,-1){20}} \put(30,30){\line(-1,1){20}}
\put(30,30){\line(-1,-1){20}} \put(3,50){$\bar\imath$}
\put(3,3){$\bar\jmath$} \put(50,3){$\bar k$} \put(51,50){$\bar l$}
\end{picture}}

\put(100,0)
 {\begin{picture}(60,60)
\put(30,30){\circle{10}} \put(20,20){\circle*{5}} \put(20,40){\circle*{5}}
\put(40,20){\circle*{5}} \put(40,40){\circle*{5}}
\put(26,26){\line(-1,-1){16}} \put(26,34){\line(-1,1){16}}
\put(34,34){\line(1,1){16}} \put(34,26){\line(1,-1){16}}
\put(3,50){$\bar\imath$} \put(3,3){$\bar\jmath$} \put(50,3){$\bar k$}
\put(51,50){$\bar l$}
\end{picture}}

\put(200,0)
 {\begin{picture}(90,60)
\put(30,30){\circle{10}} \put(60,30){\circle{10}} \put(20,20){\circle*{5}}
\put(20,40){\circle*{5}} \put(45,30){\circle*{5}} \put(70,20){\circle*{5}}
\put(70,40){\circle*{5}} \put(35,30){\line(1,0){20}}
\put(64,34){\line(1,1){16}} \put(64,26){\line(1,-1){16}}
\put(26,26){\line(-1,-1){16}} \put(26,34){\line(-1,1){16}}
\put(3,50){$\bar\imath$} \put(3,3){$\bar\jmath$} \put(80,3){$\bar k$}
\put(81,50){$\bar l$}
\put(40,34){$\bar m$}
\end{picture}}

\put(440,28){(E)}
\end{picture}
{}\vskip15mm

One indeed recognizes formulae (\ref{tim}) in the first two lines (A), (B).
Example (C) refers to the four-point function Legendre transformed only
with respect to a coupling corresponding to an operator appearing in
the intermediate state (the couplings corresponding to external legs
remain untouched). This changes the value of the four-point function
according to
\begin{equation} \label{james}
d_{ijkl} \rightarrow d_{ijkl} - c_{ijm} {1\over \Delta_m} c_{mkl}.
\end{equation}
The implications of this are quite significant. In the language of
Liouville theory, we have not modified the action, nor have we modified
any of the operators entering the four-point function. Nevertheless,
the value of the correlator changed. A similar effect was observed in
\cite{KlHa} in the course of calculating the genus one free enegy: it
changed in response to changing the dressing of any operator, even
though the Liouville action remained untouched. This suggests a
remarkable subtlety in the continuum Liouville calculations, related
perhaps to contributions from boundaries of moduli spaces.

In figure (D) we show a simpler example where a four-point
function is Legendre transformed only with respect to a coupling
corresponding to an external state, so that
\begin{equation} \label{jerry}
d_{\bar\imath jkl} =  - {1\over \Delta_i} d_{ijkl}\ .
\end{equation}
Finally, in example (E) we demonstrate the most general transformation
of the four-point function under the Legendre transform of some set of
coupling constants,
\begin{equation} \label{jim}
d_{\bar\imath\bar\jmath\bar k\bar l} =
 {d_{ijkl} \over \Delta_i\Delta_j\Delta_k\Delta_l}
- \left\{\sum_m {1\over \Delta_i} {1\over \Delta_j} c_{ijm}
 {1\over \Delta_m} c_{mkl} {1\over \Delta_k} {1\over \Delta_l}
+ \hbox{crossing terms} \right\} .
\end{equation}
The sum over $m$ runs over all intermediate operators whose
coupling constants are Legendre transformed.

The above discussion suggests that
it is useful to introduce the normalization independent quantity
\begin{equation} \label{barb}
X_{ijkl} = {1 \over \Delta_i\Delta_j\Delta_k\Delta_l}
\left\{ d_{ijkl} - {1\over2} \left(\sum_m c_{ijm} {1\over \Delta_m} c_{mkl}
+ \hbox{crossing terms} \right) \right\}^2 \ F\ .
\end{equation}
(Note the factor $1/2$ in comparing with (\ref{jim}) and (\ref{james}).)
Here, the sum over $m$ runs over {\it all} operators.
Using (\ref{theo}), (\ref{james}) and (\ref{jerry}) we establish
that $X$ is invariant under the Legendre transform with respect
to any intermediate operator, but
switches sign each time an external
operator is transformed.

Let us compare this with the behavior of correlation functions in
$c=1$ theory coupled to gravity. We introduce positively and negatively
dressed tachyon operators,
\begin{equation}
V_q={\Gamma(1+|q|)\over \Gamma(1-|q|)}
\int \d^2\sigma\, \e^{i qX +(-2+|q|)\phi}\ ,\qquad
\bar V_q={\Gamma(1-|q|)\over \Gamma(1+|q|)}
\int \d^2\sigma\, \e^{i qX +(-2-|q|)\phi}\ ,
\end{equation}
normalized to remove the usual external leg factors from the correlation
functions.
For the two-, three-, and four-point
functions of positively dressed operators in the conventional
Liouville theory we have \cite{4pt}
\begin{eqnarray}
&&\Delta_{q_i} = -{1\over \vert q_i \vert} t^{|q_i|}\ , \qquad
c_{q_1q_2q_3} =\delta(\sum_i q_i) t^{-1+(|q_1|+|q_2|+|q_3|)/2}\ ,
\nonumber \\
&&d_{q_1q_2q_3q_4} = \delta(\sum_i q_i)
t^{-2+(|q_1|+|q_2|+|q_3|+|q_4|)/2}
\left(1 -{\vert q_1+q_2\vert\over2}
-{\vert q_1+q_3\vert\over2} -{\vert q_1+q_4\vert\over2}\right).~~~
\end{eqnarray}

We may now modify the theory as in example (C), by changing the dressing
of the operator with momentum $q_1+q_2$. We assume that this flips the
sign of the Liouville energy corresponding to this state appearing
in the intermediate channel, and the resulting four-point function is
\begin{equation}
d_{q_1q_2q_3q_4} = \delta(\sum_i q_i)t^{-2+(|q_1|+|q_2|+|q_3|+|q_4|)/2}
\left(1 +{\vert q_1+q_2\vert\over2}
-{\vert q_1+q_3\vert\over2} -{\vert q_1+q_4\vert\over2}\right)\ .
\end{equation}
Similarly, the change of dressing of the $q_1+q_3$ operator
changes the sign of the corresponding term in the four-point function,
etc. It is easy to check, though, that
the quantity $X_{ijkl}$, defined
in (\ref{barb}), is invariant under these changes. In fact,
\begin{equation}
X_{q_1q_2q_3q_4}=|q_1| |q_2| |q_3| |q_4|\ .
\end{equation}
A change of dressing of one of the external operators is implemented
by $|q_i|\rightarrow -|q_i|$, which
indeed flips the sign of $X$. These properties of
$X_{q_1q_2q_3q_4}$, found with plausible assumptions about
Liouville theory, are in complete agreement with our calculations
in modified matrix model.
All other normalization independent quantities involving
three- and four-point functions can be built from (\ref{maja}) and
(\ref{barb}).  The correlators of modified operators thus agree with
the correlators of negatively dressed operators, up to possible rescalings
of the operators. We expect that this agreement extends to
higher-point functions.

It is interesting that behind the four-point function, which is
quite complex, we have uncovered a more fundamental object,
$X_{q_1q_2q_3q_4}$, which transforms very simply under changes
of the Liouville dressing. It would be interesting to check if such
simpler objects can be defined for higher-point functions.
Construction of such objects is reminiscent of the work of
Di Francesco and Kutasov \cite{DFK} who built the $c=1$ correlators in
terms of
more elementary ``vertices''. Perhaps such objects hold the key
to a better understanding of the Liouville calculations.

\section{Conclusion}

The recently improved understanding of the modified matrix models
with fine-tuned wormhole weights opens the possibility of
many new insights into random surfaces. In this paper we have
calculated some of the simplest modified correlation functions,
and there are many possible generalizations of our work.
It is remarkable that our calculations, which on a sphere reduce to
Legendre transforms, have a hidden relation to operators with the
negative branch of
Liouville dressing. We hope that more progress will come from a
deeper understanding of this effect.

\section*{Acknowledgements}

This work was supported in part by DOE grant DE-FG02-91ER40671,
NSF grant PHY90-21984,
the NSF Presidential Young Investigator Award PHY-9157482,
James S. McDonnell Foundation grant No. 91-48,
and an A. P. Sloan Foundation Research Fellowship.
C. S. is supported  by Deutsche Forschungsgemeinschaft.

\appendix
\section{\kern -1em ppendix A}

In this appendix we give a careful derivation of the fundamental
formula relating the continuum one-matrix models and
corresponding modified models:
\begin{equation} \label{uno}
\e^{\bar F(\bar t, \{t\})} = \int \d t\,\, \e^{t\bar t + F(t,\{t\})}
\end{equation}
where $\{t\} = \{t_1, t_2, \dots \}$ are couplings to scaling operators
around a particular critical point. Apparently the scaling operators
$\sigma_n$ conjugate to $t_n$, $n\ge 1$ are spectators in the
continuum formula (\ref{uno}), so that we can extend it to any
correlator without punctures
\begin{equation} \label{disconn}
\langle \bar \sigma_{i_1} \dots \bar \sigma_{i_s} \rangle_{\rm disconn.}
= \int \d t \,\, \e^{t\bar t} \langle \sigma_{i_1} \dots
\sigma_{i_s}\rangle_{\rm disconn.} \ .
\end{equation}
However, it turns out that the matrix model proof of (\ref{uno})
involves some subtleties, due to the fact that the $discrete$ scaling
operators in the old and the modified models are not exactly the same.

Let us consider a perturbation of the $k$th multicritical modified
matrix model of the form (we use even potentials for simplicity):
\begin{equation}
Z_k [\lambda, g, \{\bar \tau \}] = \int \,{\cal D}\Phi\,
\e^{-N\left(\Tr
V_k (\Phi) + (c_2 - \lambda) \Tr \Phi^4 -{g\over 2N} (\Tr \Phi^4)^2 +
\sum_{n\ge 1} \bar \tau_n \bar \sigma_n \right)} .
\end{equation}
The multicritical potential is identical to the one
of the conventional model,
\begin{equation}
V_k (\Phi) = \sum_{p\ge 1} c_p \, \Phi^{2p}\ ,
\end{equation}
and $\lambda$ is the bare cosmological constant. The bare scaling
operators have the general form
\begin{equation}
\bar \sigma_n = \sum_{p\ge 1} \bar g_{p}^{(n)} \Tr \Phi^{2p}\ .
\end{equation}
The coefficients $\bar g_{p}^{(n)}$ can be explicitly computed in this
model, because the planar limit is solvable. Let us review the main
points involved in the solution (see ref.~\cite{ABC}).

All planar properties can be extracted from the eigenvalue density. In
the one-cut phase it is given by
\begin{equation}
\rho (x) = {1\over 2\pi} \sum_{m,k\ge 0} (m+k+1)A_k (R) {\tilde
g}_{m+k+1} x^m \sqrt{R-x^2}\ ,
\end{equation}
with the definitions
\begin{eqnarray}
&& {\tilde g}_p = c_p +\bar \tau \cdot \bar g_p - g
\Bigl\langle {1\over N}\, \Tr \Phi^4 \Bigr\rangle\, \delta_{p,2}\ , \\
&& A_k (R)={2k \choose k} \left({R\over 4}\right)^k \ .
\end{eqnarray}
Since $\tilde g_p$ depends on the $\Tr \Phi^4$ condensate we are led to
a self-consistent problem given by the equation
\begin{equation}
\Bigl\langle{1\over N}\,\Tr\Phi^4 \Bigr\rangle=\int \d x\, \rho(x)\, x^4 \ .
\end{equation}
Fortunately, for the model at hand the condensate enters only linearly
in this equation, and we can solve explicitly for the feedback. The final
answer for the cosmological constant as a function of the eigenvalue
endpoint is
\begin{equation} \label{lambda}
\lambda(R) =-{1\over 2A_2^2} \left( 1+gA_2^2 - \sum_{p\ge 1} p A_p
\left(1-{p-2 \over p+2} g A_2^2\right)(c_p - c_2 + \bar \tau \cdot
\bar g_p )\right) \ .
\end{equation}
Also, the string susceptibility is given by
\begin{equation} \label{suscep}
\chi = {\d\over \d\lambda} \Bigl\langle {1\over N}\, \Tr\Phi^4\Bigr\rangle =
{A_2^2 \over 1-gA_2^2} \ .
\end{equation}
For $\bar \tau =0$ a critical point with positive susceptibility
exponent $\bar \gamma_k = {1\over k+1}$ occurs at $R_c$ when
\begin{equation} \label{critic}
\chi \sim (R-R_c)^{-1}  \qquad {\rm and} \qquad
\lambda(R) \sim \lambda(R_c) + C(R-R_c)^{k+1}\ .
\end{equation}

If we set $R_c =8$, then the critical value of $g$ is $g_c =
1/576$. Scaling operators are defined as deformations of
eq.~(\ref{critic}) which do not shift the location
of the critical point $g_c$ and $\lambda(R_c)$. This is an important
requirement because, in the end we want the couplings $\bar \tau$ to
act as sources, and non-universal dependence on $\bar\tau$ through the
value of the critical point invalidates the scaling property. As a
result, scaling operators (already in conventional matrix models)
involve one more tuning
than the multicritical potentials.

{}From eq.~(\ref{lambda}) we find the
critical conditions for the coefficients of $\bar \sigma_n$, $n\ge 1$:
\begin{equation} \label{ene}
\lambda_n (R) = {1\over 2A_2^2} \sum_{p\ge 1} pA_p \left(1-{p-2\over
p+2} gA_2^2\right) \bar g_p^{(n)} \,\, \sim \,\, (R-8)^{n+\alpha}\ ,
\end{equation}
where $\alpha$ is a positive integer. In the conventional models $g=0$
and $\alpha=0$. Remarkably, $\alpha=1$ in the modified models. Indeed,
the derivative of (\ref{ene}) is
\begin{equation} \label{deriv}
{\d\lambda_n \over \d R} = - {1-gA_2^2 \over 2A_2 R} \sum_{p\ge 1} p(p-2)
A_p \bar g_p^{(n)} \sim (R-8)^{n+\alpha-1}\ .
\end{equation}
At  the critical points with positive $\bar\gamma$ we have
 $g= 1/576$,
 $1-gA_2^2 \sim R-8$ and $\lambda_n (R)$ is at least quadratic in
$R-8$ for $n\ge 1$. This shift is ultimately responsible for the
absence of a scaling operator with dimension $\bar\gamma_k$.
{}From (\ref{deriv}) we also see that the scaling
operators in the modified model differ from those at $g=0$ only in the
coefficient of $\Tr\Phi^4$ which does not enter (\ref{deriv}), and is
determined from (\ref{ene}) by requiring stability of the critical
point. The final result is
\begin{equation} \label{dif}
\bar \sigma_n = \sigma_n + \delta g_2^{(n)}\, \Tr\Phi^4\ ,
\end{equation}
where $\sigma_n$ is the conventional bare scaling operator, and
\begin{equation} \label{gedos}
\delta g_2^{(n)} = {1\over 48} \sum_p {p(p-2)\over p+2} {2p \choose p}
2^p \bar g_p^{(n)}\ ,
\end{equation}
with a spectrum of gravitational dimensions at the $k$th critical
point:
\begin{equation}
d(\bar\sigma_n) = (n+1)\bar \gamma_k = {n+1 \over k+1}\ .
\end{equation}

We see that the bare operators in the two phases are different by a
shift of the $\Tr\Phi^4$ term. Alternatively, we can work with the same
deformations and a shifted cosmological constant $\lambda\rightarrow
\lambda + \bar\tau\cdot \delta g_2$. It turns out that the required
redefinition of $\lambda$ ensures the delicate balance of
non-universal terms needed for the scaling of formula (\ref{uno}).

Setting $\bar\tau_n = \tau_n$ we write
\begin{equation}
\bar\tau\cdot\bar\sigma = \tau\cdot\sigma + \tau\cdot\delta g_2
\Tr\Phi^4
\end{equation}
and, following ref.~\cite{KlHa} we introduce the Gaussian representation
\begin{equation} \label{gauss}
Z_k [g,\lambda,\{\tau\}] = {N\over \sqrt{2\pi g}}
\int_{-\infty}^{+\infty} \d x \, \e^{-{N^2 \over 2g}(\lambda
+\tau\cdot\delta g_2 -c_2 -x)^2} Z_k [g=0,c_2 -x, \{\tau\}]\ .
\end{equation}
Note that $Z_k$ in the integrand  is the partition function
of the $k$-th multicritical
model, deformed by the conventional scaling operators. Next we separate the
non-universal (analytic in $x$) terms on the sphere
\begin{equation}
{\rm log}Z_k [ g=0,c_2 -x,\{\tau\}] = N^2 \left( -a_1^{\{\tau\}} \,x
+{1\over 2} a_2^{\{\tau\}} \, x^2 \right) + F(x,\{\tau\}, N^2)\ .
\end{equation}
The $\tau$ dependence of the coefficients $a_1,a_2$ can be extracted
from the previously studied planar solution
\begin{equation}
a_2^{\{\tau\}} = {\d\over\d\lambda} \Bigl\langle
{1\over N}\, \Tr\Phi^4 \Bigr\rangle (g=0, \lambda= c_2) = {4 \choose 2}^2
\left({R_c \over 4}\right)^2 = 576\ .
\end{equation}
Note that $a_2 = 576$ independently of $\tau_n$. On the other hand,
$a_1$ depends linearly on $\tau$:
\begin{equation}
a_1^{\{\tau\}} = \Bigl\langle {1\over N} \Tr\Phi^4 \Bigr\rangle (g=0,
\lambda=c_2) = a_1 (\tau=0) + 576 \,\tau\cdot\delta g_2 = a_1 + a_2
\,\tau\cdot\delta g_2 \ .
\end{equation}
Finally, we  can write the Laplace transform, ready for scaling
\begin{equation}
Z_k [g,\lambda,\{\tau\}]={\rm const.}\times\int_{-\infty}^{+\infty}
\d x\, e^{N^2 a_2 x\Delta + F(x, \{\tau\}, N^2)}\ ,
\end{equation}
where $\Delta$ is given by
\begin{equation}
\Delta=\lambda-\lambda_c = \lambda-c_2 +\tau\cdot\delta g_2 -
{a_1^{\{\tau\}} \over a_2} = \lambda-c_2 - {a_1\over a_2}\ ,
\end{equation}
and the scaling variable $\Delta$ is independent of the scaling
deformations $\tau_n$. As a result, we can safely take derivatives in
$\tau_n$ to obtain a discrete version of (\ref{disconn}). We see that
the slight difference between $\bar\sigma_n$ and $\sigma_n$ ensures
the stability of the critical point $\lambda_c$ under such
deformations. Formula (\ref{uno}) follows then from the scaling
\begin{equation}
x\sim t \, N^{2\over \gamma-2}\ ,  \qquad
\Delta\sim \bar t  \,N^{2\over\bar\gamma-2}\ , \qquad
\tau_n \sim t_n \, N^{2\gamma n \over \gamma-2}
\end{equation}
with ${1\over 2-\gamma} + {1\over 2-\bar\gamma} =1$.

\end{document}